\documentclass[fleqn,10pt]{wlscirep}

\title{Analog simulator of integro-differential equations with classical memristors}
\author[1,2,3,$\dag$]{G. Alvarado Barrios}
\author[1,2]{J. C. Retamal}
\author[3,4,5]{E. Solano}
\author[4,$\dag \dag$]{M. Sanz}

\affil[1]{Departamento de F\'isica, Universidad de Santiago de Chile (USACH), 
	Avenida Ecuador 3493, 9170124, Santiago, Chile}
\affil[2] {Center for the Development of Nanoscience and Nanotechnology, Santiago, Chile}
\affil[3] {International Center of Quantum Artificial Intelligence for Science and Technology (QuArtist) \\ and Department of Physics, Shanghai University, 200444 Shanghai, China}
\affil[4] {Department of Physical Chemistry, University of the Basque Country UPV/EHU,  Bilbao, Spain}
\affil[5] {IKERBASQUE, Basque Foundation for Science, Maria Diaz de Haro 3, 48013 Bilbao, Spain}
\affil[$\dag$]{gabriel.alvarado@usach.cl}
\affil[$\dag \dag$]{mikel.sanz@ehu.eus}

\begin{abstract}
An analog computer makes use of continuously changeable quantities of a system, such as its electrical, mechanical, or hydraulic properties, to solve a given problem. While these devices are usually computationally more powerful than their digital counterparts, they suffer from analog noise which does not allow for error control. We will focus on analog computers based on active electrical networks comprised of resistors, capacitors, and operational amplifiers which are capable of simulating any linear ordinary differential equation. However, the class of nonlinear dynamics they can solve is limited. In this work, by adding memristors to the electrical network, we show that the analog computer can simulate a large variety of linear and nonlinear integro-differential equations by carefully choosing the conductance and the dynamics of the memristor state variable. We study the performance of these analog computers by simulating integro-differential models related to fluid dynamics, nonlinear Volterra equations for population growth, and quantum models describing non-Markovian memory effects, among others. Finally, we perform stability tests by considering imperfect analog components, obtaining robust solutions with up to $13\%$ relative error for relevant timescales.
\end{abstract}

\begin{document}
\maketitle


\section*{Introduction}
Analog computers employ continuously tunable quantities of a system, such as its electrical, mechanical, or hydraulic properties, to codify and solve a given problem. Usually, the dynamics of an analog computer perfectly matches the dynamics of the simulated system. The advantage of analog computers lies on the computational power provided by real-time operation and complete parallelism, so that they require less resources than a digital counterpart for the simulation. On the other hand, the accuracy of an analog computer is limited by its computing elements and by the quality of its analog components.
Despite the astonishing evolution of digital computers during the last decades, the interest in analog computers has re-emerged in recent years \cite{MacLennan2014}. For example, posible applications as math co-processors in VLSI architectures \cite{Cowan2005,Cowan2006}, and wave-based analog computation in metamaterials have recently been proposed \cite{Silva2014metamaterials,AbdollahRamezani2015,Youssefi2016}. 

An electrical analog computer is an active network composed of electrical elements, namely, resistors, capacitors and operational amplifiers which, connected together, are capable of simulating any set of linear ordinary differential equations~\cite{Kodali1967}. In this case, the solutions of these equations are encoded into the time evolution of the voltage waveform produced by the analog computer. Indeed, electrical analog devices can only work with a single independent variable when codified in time \cite{Pour-El1974,Rubel1988}. Nonetheless, by making use of finite-difference methods, it is also possible to solve partial differential equations \cite{Kodali1967,Ratier2012}.

Resistive memory effects, in which the resistance depends on the history of charges crossing through a material, can be described by Kubo's response theory \cite{Kubo1957}, and have been observed as resistance switching phenomena in thin films \cite{Simmons1967,Argall1968}. The connection between resistance switching devices and memristive devices \cite{Chua1971memristor,Chua1976memristive} was first made in 2008 by HP Labs \cite{Strukov2008themissing}. Afterwards, various other physical implementations of memristors have been reported \cite{Jo2010nanoscale,Chang2011,Chanthbouala2012ferroelectric,Kim2012ferroelectric}. The memory and non-volatility properties of the memristor \cite{Chua2011resistance} have sparked great interest in its applications in the fabrication of digital memories \cite{Waser2009redoxbased}, performance of computational tasks \cite{Borghetti2010memristive,Pershin2012Neuromorphic,Yang2012memristive,Kulkarni2012,Cassinerio2013} and recently, in the design of a perceptron \cite{Silva2018}, which motivates the possibility of building memristor-based hardware for a physical neural network and proved their universality as approximators. While most of the interest in memristors lies in applications in digital computation, there have also been proposals of utilization of memristors in programmable analog components of electrical networks \cite{Pershin2010practical} and in the design of highly efficient operational amplifiers \cite{Jahromi2017ultralow}. Regarding the solution of differential equations, memristor arrays have been employed as accelerators for digital solvers \cite{Richter2015, Gallo2018} and, recently, a digital approach based on memristor crossbar arrays have been employed to solve partial differential equations \cite{Zidan2018} with high precision. Nonetheless, a natural question is whether the characteristic properties of memristors could extend the applications of electrical analog computers. 

In this work, by adding classical memristors to the network of an electrical analog computer, we show that it can simulate a large variety of linear and nonlinear integro-differential equations with interest in mathematics and physics. This is possible by appropriately choosing the conductance of the memristors and the dynamics of its state variable. We study the performance of these analog computers by simulating integro-differential models of fluid dynamics, nonlinear Volterra equations for population growth, and quantum models describing non-Markovian memory effects, among others. Finally, by considering $10 \%$ error in the analog components, we perform stability tests of the dynamics, showing the robustness of the simulations with up to $13\%$ error for relevant timescales. It is noteworthy to mention that the operation conditions of an off-the-shelf memristor fit those of usual circuit toolboxes.

\section*{Electric Analog Computer}

To simulate a linear ordinary differential equation, the analog computer only requires the following operations: (i) summation, (ii) sign inversion, (iii) integration and (iv) multiplication by a constant.  

\begin{figure}[t!]
	\center
	\includegraphics[width=0.7\linewidth]{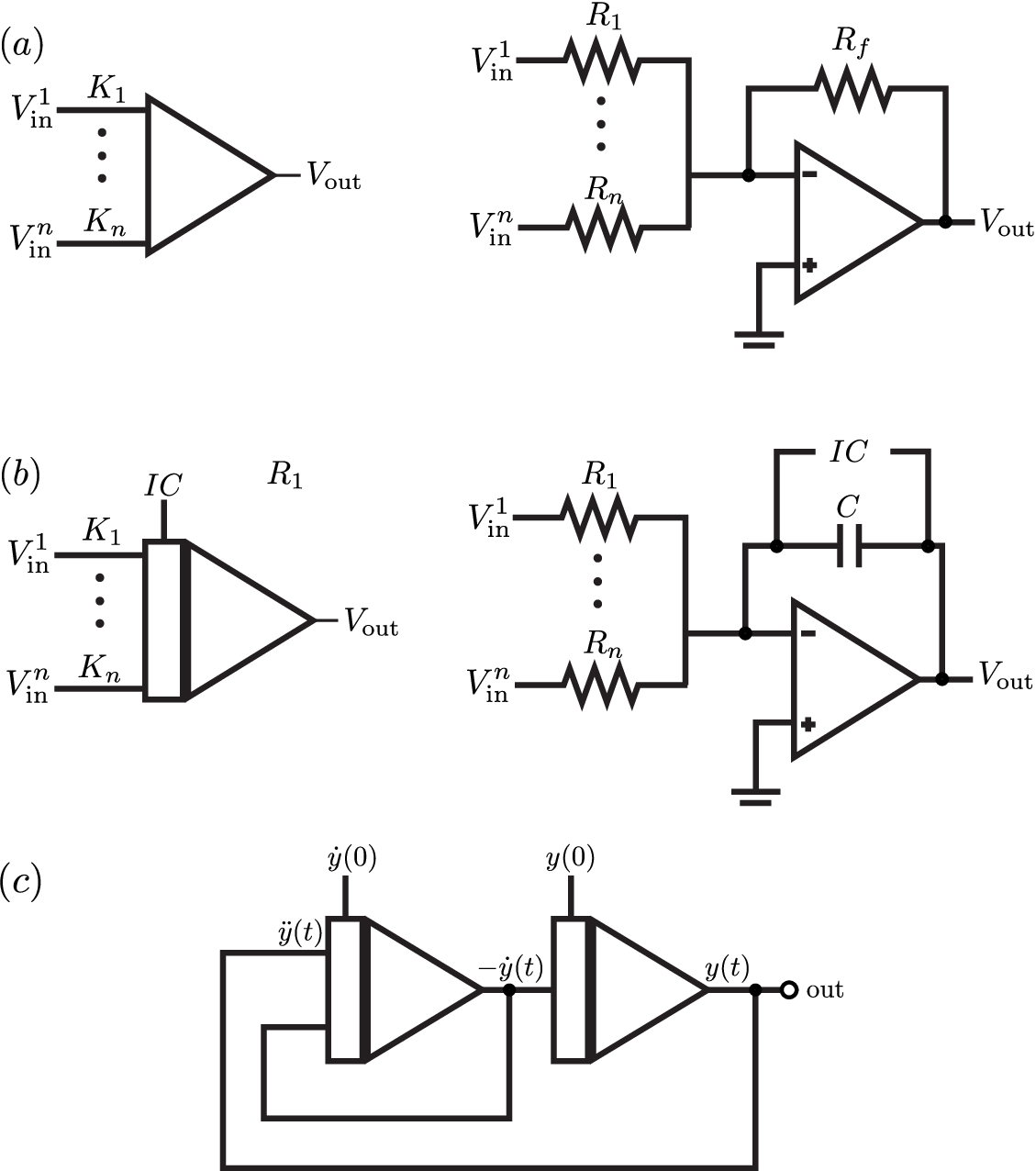}
	\vspace{.2cm}
	\caption{Symbolic representation and corresponding circuit of the computing elements of an analog computer.  (a) Adder for $n$ inputs. (b) Integrator for $n$ inputs. After the initial charge of the capacitor, which is used to set the initial conditions, that part of the circuit is switched off from the rest. (c) Circuit diagram for the simulation of the second-order linear ordinary differential equation of Eq.~(\ref{example}).}	
	\label{Circuit_elements}
\end{figure}

We will describe briefly the implementation of the aforementioned operations for the analog computer. First, let us consider the summation and sign inversion operations, which are performed by an adder circuit whose transfer characteristics for $n$ inputs are given by
\begin{equation}
V_{\text{out}} = - \sum_{i}^{n} K_{i}V_{\text{in}}^{i},
\end{equation} 
where $V_{\text{out}}$ is the output voltage and ${K_{i}=R_{f}/R_{i}}$. The symbolic representation and equivalent circuit of the adder is shown in \textbf{Fig.~\ref{Circuit_elements}(a)}. Sign inversion can be implemented by an adder with a single input with ${K_{1} = 1}$.
Next, integration, is meant to be the most important operation in an analog computer. It can be implemented by an integrator circuit, whose transfer characteristics for $n$ inputs are given by 
\begin{equation}
V_{\text{out}}(t) = - \frac{1}{C} \int_{0}^{t} \sum_{i} \frac{V_{\text{in}}^{i}(\tau)}{R_{i}} d\tau + IC,  \label{integrator}
\end{equation}
where the constant $IC$ stands for the initial condition. The symbolic representation and equivalent circuit of the integrator is shown in \textbf{Fig.~\ref{Circuit_elements}(b)}. For the adder and integrator circuits, we consider ideal operational amplifiers.

The last linear operation, multiplication by a constant, can be implemented with a potentiometer, which is described by
\begin{equation}
V_{\text{out}} = \alpha V_{\text{in}},
\end{equation}
where $\alpha<1$ is a constant that characterizes the device.

In addition, analog computers can also simulate nonlinear dynamics to some extent, by considering signal generators and function multipliers, which enable the mathematical multiplication of two signals. 

We will consider a simple example to illustrate the configuration of an analog computer for the simulation of a linear ordinary differential equation. Let us consider the following second-order linear differential equation
\begin{equation}
\ddot{y}(t) = -\dot{y}(t) + y(t). \label{example}
\end{equation}
This equation is written in this form to remark that $\ddot{y}$ can be obtained as the sum of $-\dot{y}$ and $y$, which are obtained from integrating $\ddot{y}$ once and twice, respectively. The computer diagram that implements Eq.~(\ref{example}) is shown in \textbf{Fig.~\ref{Circuit_elements}}(c). Therefore, we see that integrator circuits enable the solution of ordinary differential equations by making use of feedback connections, which make the input and output voltage in Eq.~(\ref{integrator}) to be equal. In addition, several integrators connected in series stand for subsequent integrations of the desired variable. In this way, the simulation of a linear differential equation of order $n$ would require $n$ integrators.

This method can be extended to a system of $m$ linear ordinary differential equations of order $n$, which can be written as

\begin{eqnarray}
\label{diffeq} 
\sum_{\ell=1}^{m}\sum_{k=0}^{n} a_{\ell,k}\frac{d^{(k)} y_{\ell}(t)}{dt^{k}} = 0.   \label{generalcase}
\end{eqnarray}
The simulation of such a system requires $nm$ integrators, and since the summation can be absorbed into the integrators, it requires at most $\frac{nm}{2}$ sign inverters. 

It is important to note that, in order to solve a differential equation with an analog computer, it is not necessary to know the voltage waveform at any point of the electrical network, provided that the circuit was properly constructed, the desired solution is obtained by measuring the voltage signal at the output of the circuit. We remark that analog computers are best suited for solving systems of ordinary differential equations, and aditionally, it is possible to use them to solve partial differential equations by making use of finite differences approximations \cite{Kodali1967}. Since analog computers are better suited to solve time-dependent problems, the non-temporal variable is discretized to obtain a system of nonlinear ordinary differential equations for the time variable.

\section*{Memristive analog computation}
A voltage-controlled memristor can be described by the following current-voltage relation \cite{Chua1971memristor}
\begin{eqnarray}
i &=& g(\omega)\, v, \\  \label{idealmemristor}
\dot{\omega} &=& f(v),
\end{eqnarray}
where $i(t)$ and $v(t)$ are the current and voltage across the device, respectively. The quantity $\omega$ is an internal state variable specified by the physical implementation of the memristor. For ideal voltage-controlled memristors the state variable corresponds to the magnetic flux $\phi$, which means $f(v)=v$. The quantity $g(\omega)$ is the so-called memductance which has units of conductance and is a function of the state variable. Since the memristor is a passive circuit element, the memductance $g(\omega)$ must always be positive.

The notion of memristive behavior can be extended to a broader class of system whose characteristics resemble those of the memristor, and are called memristive systems \cite{Chua1976memristive}. These are defined by
\begin{eqnarray}
i &=& g(\boldsymbol{\omega},v,t)\,v, \label{memcurrent} \\ 
\dot{\boldsymbol{\omega}} &=& \mathbf{f}(\boldsymbol{\omega},v,t).     \label{meminternal}
\end{eqnarray}
Where in general, $\boldsymbol{\omega}$ represents a set of $n$ state variables, and $\mathbf{f}$ and $g$ are continuous functions.

\begin{figure}[t!]
	\center
	\includegraphics[width=.35\linewidth]{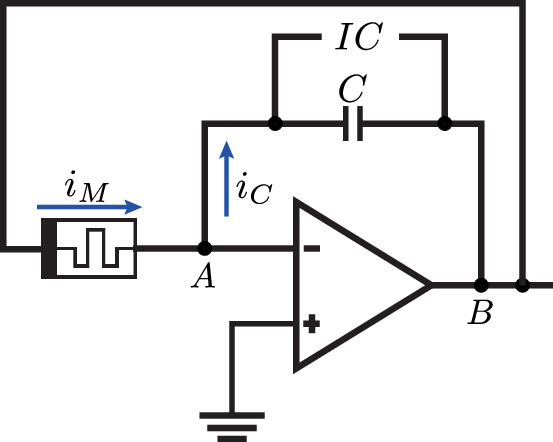}
	\vspace{.2cm}
	\caption{Integrator circuit in which we have replaced the resistor for a memristor. After the initial charge of the capacitor, the part denoted by $IC$ is switched off from the rest of the circuit.}
	\label{MemIntegrator}	
\end{figure}

Now, to study the inclusion of memristive devices into an analog computer we will consider the case of an integrator circuit with a single input, where we substitute the resistor for a memristor, as is shown in \textbf{Fig.~\ref{MemIntegrator}}. We will consider that the memristor has a single state variable $\omega$. In the figure, the current across the memristor, $i_{M}$, is given by Eq.~(\ref{memcurrent}). Then, by Kirchhoff's law at node $A$, we have 

\begin{eqnarray}
i_{M} &=& i_{C}, \\
g(\omega,v,t)\, v &=& - C\frac{dv}{dt}. \label{mem-int}
\end{eqnarray}
Integrating Eq.~(\ref{meminternal}) allows to write Eq.~(\ref{mem-int}) as
\begin{equation}
\dot{v}(t) = - \frac{1}{C}g \bigg(\int_{0}^{t} f(\omega,v,\tau)d\tau \boldsymbol{,}\, v(t) \boldsymbol{,}\, t \bigg) v(t), \label{memintegrator}
\end{equation}
which in general will be a nonlinear integro-differential equation that will be specified by $g$ and $f$, which describe the memductance and the memristor state variable dynamics, respectively. It must be noticed that these functions are specified by memristor engineering.

We now show the type of equations that can be solved using this circuit. Let us consider the equation that Volterra introduced for single populations in the study of growth of biological populations \cite{Volterra2005}, 
\begin{equation}
\dot{N}(t) = N(t)\bigg[a - bN(t) - \int_{0}^{t} k(t-s)N(s) ds \bigg], \quad  t \in \boldsymbol{R}^{+},
\end{equation} 
in which $N(t)$ is the population size at time t, $a$ and $b$ are positive rate constants, and $k(t)$ is the ``hereditary'' influence. 

Now, in order to simulate this equation, we consider a memristive system as described by Eq.~(\ref{memcurrent}) and Eq.~(\ref{meminternal}). By assuming ${g(\omega,v,t) = -a + bv + k_{1}(t)\omega}$ and $\dot{\omega} = k_{2}(t)v$, then, substituting into Eq.~(\ref{memintegrator}) we have
\begin{equation}
\dot{v}(t) = \bigg(a - bv(t) - \int_{0}^{t}k(t,\tau)v(\tau)d\tau\bigg) v(t), \label{VolterraG}
\end{equation}
where we have chosen $C=1$ and $k(t,\tau)=k_{1}(t)k_{2}(\tau)$. This is restricted, however, to kernels that can be separated in this way, otherwise it is only possible to simulate single-variable kernels. This equation can be implemented with a single integrator and memristor as shown in \textbf{Fig.~\ref{MemIntegrator}}.

By connecting integrator circuits in series, we can generate nonlinear integro-differential equations of higher-order. For example, if we consider $n$ integrators in series, where only the first integrator circuit contains a memristor, then we can generate the $nth$-order integro-differential equation
\begin{equation}
\frac{d^{(n)}v}{dt^{(n)}} = - \frac{1}{C}g \bigg(\int_{0}^{t} f(\omega,v,\tau)d\tau \boldsymbol{,}\, v(t) \boldsymbol{,}\, t \bigg) \, v(t).
\end{equation}
On the other hand, if each of the $n$ integrators contains a memristor, then the $nth$-order integro-differential equation will involve the composition of $n$ memductance functions $g_{1},\dots,g_{n}$ corresponding to each memristor. The resulting equation can be written as follows
\begin{flalign}
&\frac{d^{(n)}v}{dt^{(n)}} = \\ \nonumber
&- \frac{1}{C} g_{n}\bigg(g_{n-1} \bigg(\dots g_{1} \bigg(\int_{0}^{t} f(\omega,v,\tau)d\tau \boldsymbol{,}\, v(t) \boldsymbol{,}\, t \bigg) \dots \bigg) \bigg)\, v(t).
\end{flalign}
This allows for the simulation of a wide class of nonlinear integro-differential equation.

It is also possible to solve first-order linear integro-differential equations with a suitable change of variables. If we consider Eq.~(\ref{memintegrator}) and define the memristor by ${g(\omega,v,t) = -\omega}$ and $\dot{\omega} = k(t)\ln(v)$, we obtain 
\begin{equation}
\dot{v}(t) = \frac{1}{C} \bigg(\int_{0}^{t} k(s)\ln(v(s)) ds\bigg)  v(t).
\end{equation}
Then, with $u = \ln(v)$, we have
\begin{equation}
\dot{u}(t) =  \frac{1}{C} \bigg(\int_{0}^{t} k(s)u(s) ds\bigg). 
\end{equation}
Notice that we can avoid the sign inversion of the integrand  by normalizing the voltage such that $0\leq v \leq 1$. In this way, it is possible to simulate several linear integro-differential equations coming from quantum models. An example of this is a Volterra equation describing non-Markovian quantum memory effects \cite{Rodriguez2017}, given by 	
\begin{figure}[t!]
	\center
	\includegraphics[width=.4\linewidth]{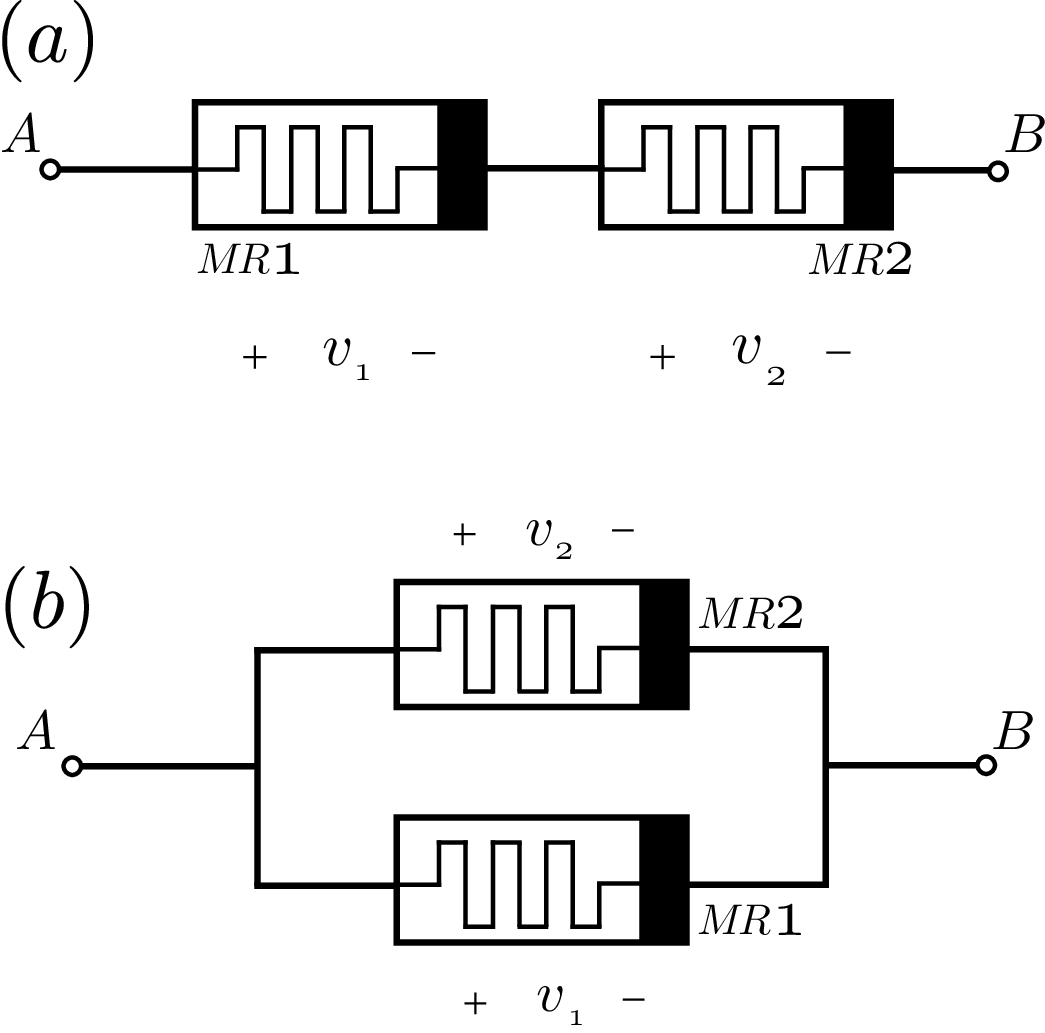}
	\caption{Composite memristor circuit (a) connected in series and (b) connected in parallel with same polarities. Here, $v_{1}$ and $v_{2}$ are the magnetic voltage of the first and second memristor, respectively.}
	\label{Compositemem}
\end{figure}
\begin{equation}
\partial_{t}\rho = \int_{0}^{t} K(t,s) \mathcal{L} \rho.
\end{equation}
Since $\mathcal{L}$ is a linear superoperator, the integral kernel will involve linear terms in the elements of $\rho$. This leads to a system of integro-differential equations for the elements of $\rho$ which can be implemented by appropriately choosing the dynamics of the internal state variable of the memristor.

Aditionally, linear integro-differential equations that appear in the model of turbulent diffusion \cite{Velikson1975}, have the form
\begin{equation}
\dot{u}(t) + p(t)u(t) + \int_{0}^{t}K(t,s)u(t-s)u(s)ds = 0. \label{turbulent}
\end{equation}
Equations of this type can be implemented with a memristor defined by $g(\omega,v,t) = \alpha(t)k_{1}(t)\omega$ and $\dot{\omega} = \frac{k_{2}(t)}{\alpha(t)^{2}} \ln(v)^{2}$, such that $K(t,s) = k_{1}(t)k_{2}(s)$, and where ${\alpha(t) = \exp\bigg(\int_{0}^{t} p(s)ds\bigg)}$. By using ${z(t) = \alpha(t)u(t)}$, we can configure Eq.~(\ref{turbulent}) into the analog computer as the following equation, 
\begin{equation}
\dot{z}(t) =  - \bigg(\int_{0}^{t} k'(s)z(s)^{2} ds\bigg),  \label{linearInt}
\end{equation}
with $k'(s) = \frac{k(s)}{\alpha(s)^{2}}$. Next, connecting the output $z(t)$ to a signal multiplier with the signal $\frac{1}{\alpha(t)}$, we recover the solution $u(t)$ of Eq.~(\ref{turbulent}).
\begin{figure}[t!]
	\center
	\includegraphics[width=.4\linewidth]{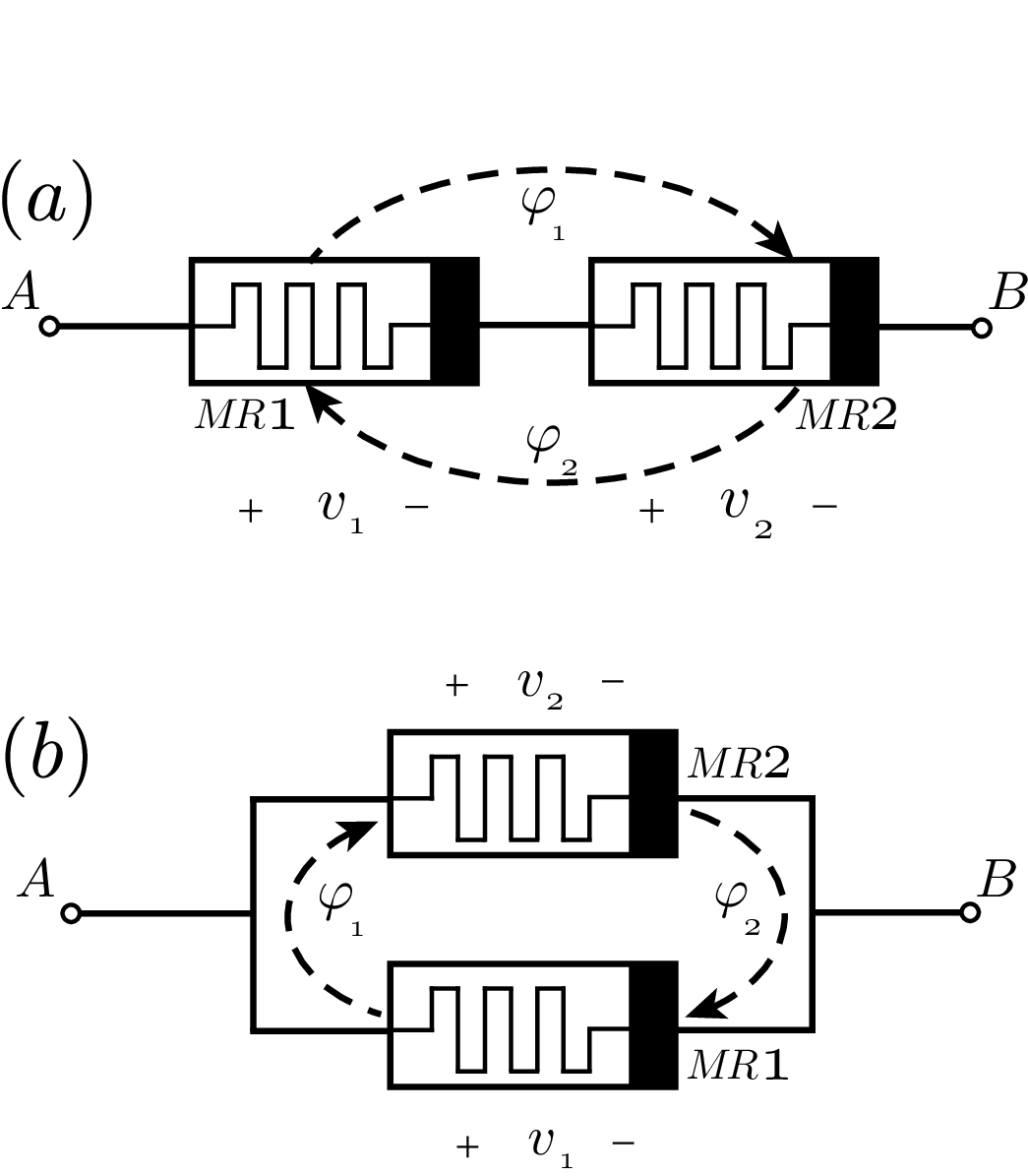}
	\caption{Coupled memristors with flux interaction, (a) connected in series and (b) connected in parallel with same polarities. Here,  $\phi_{1}$ $(\phi_{2})$, $v_{1}$$(v_{2})$ are the magnetic flux and voltage of the first (second) memristor, respectively. When the timescale of the state variable dynamics is dominant over other timescales of each memristor it is possible to describe the composite memristor circuit as an equivalent memristor.}
	\label{Coupledmem}
\end{figure}

\section*{Equivalent memristors for enhancing simulations}
We have seen that the integro-differential equation that can be implemented with the integrator circuit of \textbf{Fig.~\ref{MemIntegrator}} depends on the memductance, $g$, and internal variable dynamics, $f$, in Eqs.~(\ref{memcurrent}) and  (\ref{meminternal}), respectively. However, the choice of the memductance function $g$ is restricted by currently available memristors. A possible way for broadening the class of available memductance functions is to consider a composite memristor circuit, connected in series or in parallel \cite{Budhathoki2013, Hu2015}.

The behavior of composite memristor circuits is nontrivial, it involves transient phenomena which depend on the initial states of each memristor and is also affected by their polarities. Nonetheless, it has been shown that, for ideal voltage-dependent memristors, a composite memristor circuit involving either in parallel or in series connection can obey the same rules as with standard resistors \cite{Budhathoki2013, Hu2015}. This behavior depends on the memristor internal relaxation time and on the state variable dynamics. Assuming that the timescale of memristor relaxation is much shorter than the timescale associated with the state variable dynamics, then we can neglect the former and ignore transient phenomena.

We can expect that, under the aforementioned assumptions on timescales, a composite memristor circuit involving memristive systems will obey a dynamical model similar to Eqs.~(\ref{memcurrent}) and  (\ref{meminternal}). Then, we look to describe the composite memristor circuit as an equivalent memristor with an equivalent memductance. For a composite memristor circuit connected in series, as shown in \textbf{Fig.~\ref{Compositemem}} (a), the equivalent memductance is given by 
\begin{equation}
\frac{1}{g_{\text{equiv}}(\omega,v,t)} = \frac{1}{g_{1}(\omega_{1},v_{1},t)} + \frac{1}{g_{2}(\omega_{2},v_{2},t)},
\end{equation}
whereas for memristors connected in parallel, it holds 
\begin{equation}
g_{\text{equiv}}(\omega,v,t) = g_{1}(\omega_{1},v_{1},t) + g_{2}(\omega_{2},v_{2},t),
\end{equation}
with $v = v_{1} + v_{2}$, and $\omega = \omega_{1} + \omega_{2}$. Then, it is possible to approximate a memductance function, $g(\omega,v,t)$, by considering its Taylor expansion, $g(\omega,v,t) = c_{1} + c_{2} v + \frac{1}{2!} c_{3} v^{2} + ...$, and construct a  composite memristor circuit connected in parallel such that each memristor in the circuit provides one of the terms of the desired Taylor series. Then, the equivalent memductance, given by 
\begin{equation}
g_{\text{equiv}}(\omega,v,t) = g(\omega_{1},v_{1},t) + g(\omega_{2},v_{2},t) + g(\omega_{3},v_{3},t) + ... \label{Taylor}
\end{equation} 
is an approximation of the desired memductance function, $g(\omega,v,t)$, provided that $g(\omega_{1},v_{1},t) = c_{1}$,  $g(\omega_{2},v_{2},t) = c_{2} v$,  $g(\omega_{3},v_{3},t) = \frac{1}{2!} c_{3} v^{2}$ and so on. In this way, the desired memductance function, $g(\omega,v,t)$, can be approximated with a controllable accuracy which depends on the amount of terms of the Taylor series implemented.

In addition, when memristors are set in proximity, they may mutually interact \cite{Yu2015,Cai2014}, affecting the dynamics of their internal variables, and consequently, their memductances. This leads to a different class of integro-differental equations that can be implemented in the analog computer by including coupled memristors into the network. Let us consider an ideal voltage-controlled memristor system described by 
\begin{eqnarray}
i_{1}(t) &=& g_{1}(\phi_{1},\phi_{2})v_{1}(t), \\
i_{2}(t) &=& g_{2}(\phi_{1},\phi_{2})v_{2}(t), \\
\frac{d\phi_{1}}{dt} &=&  v_{1}(t), \, \frac{d\phi_{2}}{dt} = v_{2}(t),
\end{eqnarray}
where we consider $g_{1}(\phi) = \alpha \phi + \beta $, and $\alpha,\beta$ are constants. Then, we have 
\begin{eqnarray}
g_{1}(\phi_{1},\phi_{2}) = \alpha_{1}\phi_{1} + \beta_{1} + \kappa_{2}\phi_{2}, \\
g_{2}(\phi_{1},\phi_{2}) = \alpha_{2}\phi_{2} + \beta_{2} + \kappa_{1}\phi_{1}. 
\end{eqnarray}
We can consider again two coupling cases, memristors connected in series and in parallel. For two memristors connected in series, as shown in \textbf{Fig.~\ref{Coupledmem}(a)}, we define $v_{12}= v_{1} + v_{2}$, and $\phi_{12}= \phi_{1} + \phi_{2}$. By applying Kirchhoff's voltage law, together with the constitutive relations of the memristors, one obtains a set of coupled differential equations for the internal variables of the memristors \cite{Yu2015}. In the special case of $\alpha_{1} = \alpha_{2} = \alpha$, $\beta_{1} = \beta_{2} = \beta$, $\kappa_{1} = \kappa_{2} = \kappa$, the equivalent memductance, defined by $i(t) = g_{12}(\phi_{12}) v_{12}$, is consequently given by (see Ref. \cite{Yu2015})
\begin{equation}
g_{12}(\phi_{12}) = \frac{1}{2}\frac{\alpha^{2}\phi_{12}^{2} + 2\alpha \beta \phi_{12}}{4\beta + 2\alpha \phi_{12}} + \beta. 
\end{equation} 
This result may allow us to implement an additional kind of integro-differential equation, related to non-Markovian dynamics, given by the following expression:
\begin{equation}
C\dot{v}_{12} = \bigg( \frac{1}{2}\frac{\alpha^{2}(\int_0^{t} v_{12}(\tau)d\tau)^{2} + 2\alpha \beta (\int_0^{t} v_{12}(\tau)d\tau)}{4\beta + 2\alpha (\int_0^{t} v_{12}(\tau)d\tau)} + \beta\bigg) v_{12}. 
\end{equation} 
On the other hand, for coupled memristors connected in parallel, as shown in \textbf{Fig.~\ref{Coupledmem}(b)}, we have $i = i_{1} + i_{2}$, and $\phi_{12}= \phi_{1} + \phi_{2}$. In this case the total memducante of the dual coupled memristors in parallel connection is given by the expression (see Ref.~\cite{Yu2015})
\begin{equation}
g_{12}(\phi_{12}) = (\alpha_{1} + \alpha_{2} + \kappa_{1} + \kappa_{2})\phi_{12} + \beta_{1} + \beta_{2}.
\end{equation} 
This is, the memristors coupled in parallel operate as a new memristor, and the equivalent memductance is obtained as the sum of the individual memductances. We see that, in this case, coupled memristors connected in parallel behave very similar to the uncoupled case. The main difference is the appearance of additional parameters.\\
We want to mention that to consider a large memristor array in our case could benefit the proposal, specially in obtaining a good approximation of any desired memductance function by constructing its Taylor expansion with an equivalent memristor array connected in parallel. The main limitation of using a large number of memristors lies in the error accumulation due to imperfections in the devices, as will be seen later. In this sense, the flexibility with large arrays increases as the device fabrication improves.

\section*{Adaptability with other classes of memdevices}
Another possibility for broadening the class of integro-differential equations that can be implemented by the analog simulator is to consider other classes of memdevices, such as memcapacitive and meminductive devices \cite{DiVentra2009}. Here we will consider the substitution of the capacitor in the integrator circuit of Fig.~\ref{MemIntegrator} for a voltage-controlled memcapacitive device which can be specified as follows: 

\begin{eqnarray}
q &=& C(\boldsymbol{x},v,t)\,v, \label{memcharge} \\ 
\dot{\boldsymbol{x}} &=& \mathbf{\phi}(\boldsymbol{x},v,t),     \label{capinternal}
\end{eqnarray}

where $q$, $v$, $x$ and $C$,  are the charge, voltage, internal variable and memcapacitance in the memcapacitive system, respectively. Then, performing the same calculation in node $A$ of Fig.~\ref{MemIntegrator} with Kirchhoff's law, as in Eq.(\ref{mem-int}), we obtain

\begin{eqnarray}
i_{M} &=& i_{C}, \\
g(\omega,v,t)\, v &=& - \big( \frac{dC}{dt}v(t) + C(x,v,t)\frac{dv}{dt} \big) . \label{mem-capint}
\end{eqnarray}
his leads to the following integro-differential equation
\begin{equation}
\dot{v}(t) = - \frac{1}{C(x,v,t)}g \bigg(\int_{0}^{t} f(\omega,v,\tau)d\tau \boldsymbol{,}\, v(t) \boldsymbol{,}\, t \bigg) v(t) - \frac{1}{C(x,v,t)} \frac{dC}{dt} v(t) , \label{capmemint}
\end{equation}
This naturally adds a new parameter, the memcapacitance $C(x,v,t)$, that increases the flexibility in the class of integro-differential equations that can be implemented with the analog simulator. For instance, we can see that by considering $C(x,v,t) = x$ \cite{DiVentra2009} we can include a linear term in the voltage that will depend on the dynamics of the internal variable of the memcapacitive system. This can ease the requirements on the memductance which for some applications can become a very intricate function of the internal variable, voltage and time. 
\section*{Numerical robustness}    
\begin{figure}[t!]
	\center
	\includegraphics[width=.6\linewidth]{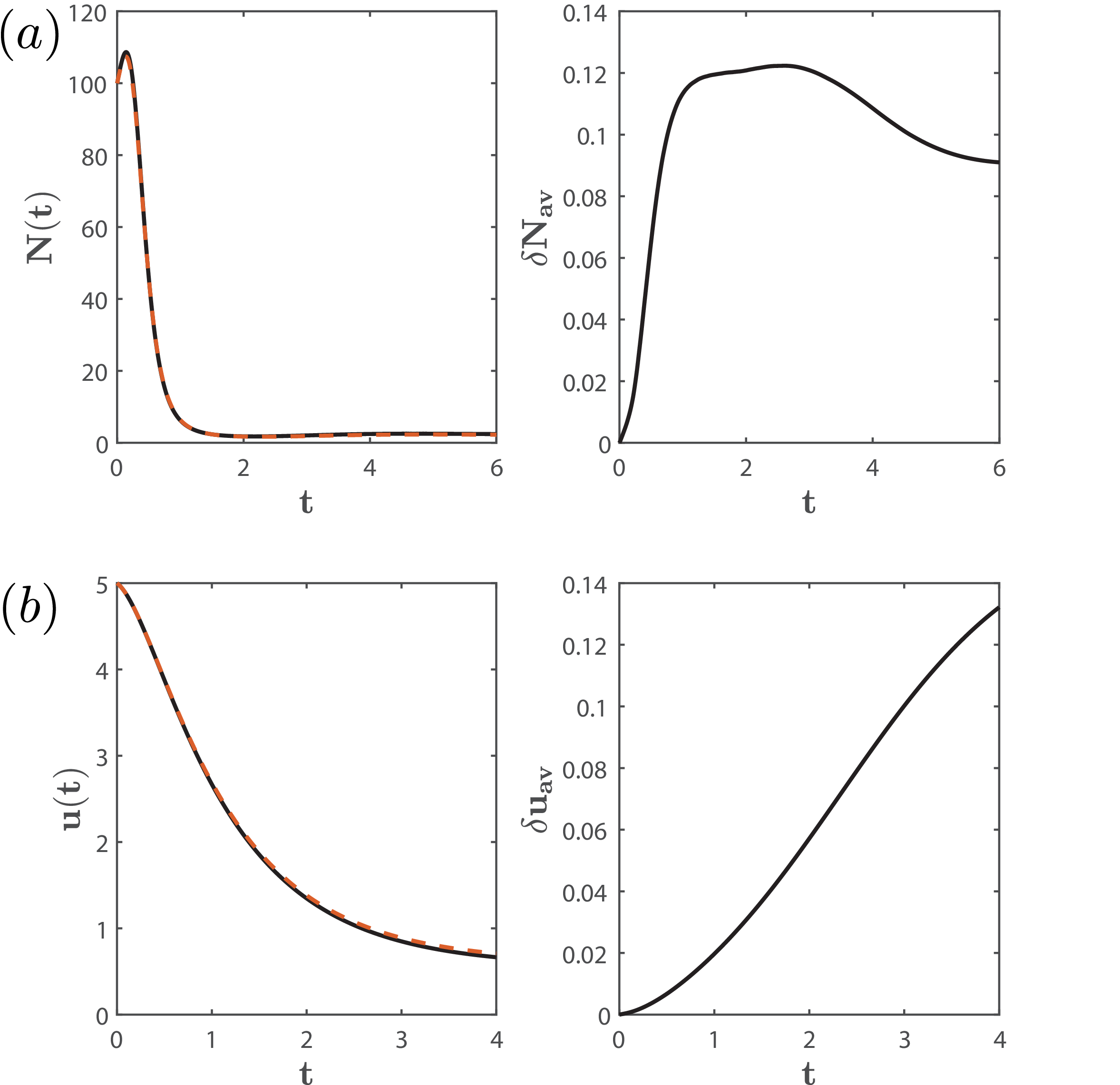}
	\caption{Stability test of the solution of (a) Eq.~(\ref{Exampleeq1}) and (b) Eq.~(\ref{Exampleeq2}). (Left) Shows the exact solution (black solid line) and the solution considering imperfect analog components (red dashed line) with up to 10\% error in their corresponding coefficients. (Right) Shows the average relative error over $100$ iterations of the simulation with imperfect analog components.}
	\label{Stability_VandT}
\end{figure}

In order to provide an example of the previous procedures and to test the stability of the solutions, let us consider the following integro-differential equation 
\begin{eqnarray}
\dot{N}(t) &=& 2N(t) - 0.001N(t)^{2} \nonumber \\
&-& N(t)\int_{0}^{t} \frac{s}{1+s}e^{-(t-s)}N(s) ds.  \label{Exampleeq1}
\end{eqnarray}
This equation can be implemented by taking $C = 1$ and chosing $g(\omega,v,t) = -2 + 0.001v + e^{-t}\omega$ and $\dot{\omega}=e^{s}\frac{s}{1+s} v$ in Eq.~(\ref{memintegrator}). In \textbf{Fig.~\ref{Stability_VandT}(a)}, we show the exact solution of this equation compared against the case with up to $10\%$ error in the coefficients of Eq.~(\ref{Exampleeq1}), which stands for imperfections in the analog components of the computer. We can see that the average relative error stabilizes at $\sim 10\%$ indicating that the solution is robust against imperfections in the analog components.
As a second example we consider the following integro-differential equation	
\begin{equation}
\dot{u}(t) = -\bigg( \frac{1}{8}e^{-2t}u(t) + \int_{0}^{t}\frac{1}{2} e^{-(t+s)}u(s)^{2}ds \bigg).  \label{Exampleeq2}
\end{equation}
This equation can be taken into the form of Eq.~(\ref{linearInt}) by considering $\alpha(t) = \exp(-1/16(e^{-2t - 1}))$,  $k_{1}(t) = 1/2 e^{-t}$, $k_{2}(s) =  e^{-s}$ and ${\dot{\omega} = k_{2}(s) \ln(v)^{2}}$. 
In \textbf{Fig.~\ref{Stability_VandT}(b)}, we show the solution of this equation compared against the case with up to $10\%$ error in the coefficients of Eq.~(\ref{Exampleeq2}), as well. We observe that, in this case, the error increases continuously, as it is usually the case in analog computers, reaching up to $13\%$ within the timescale considered.

\section*{Discussion}

We have studied the inclusion of memristors into the network of electric analog computers. We have found that this addition enables analog computers to simulate linear and nonlinear integro-differential equations by appropriately choosing the memductance and the dynamics of the memristor state variable. This broadens the applicability of analog computers which, otherwise, could only solve systems of ordinary differential equations. Additionally, we have numerically studied the performance of these analog computers by simulating integro-differential models related to fluid dynamics, nonlinear Volterra equations for population growth, and quantum models describing non-Markovian memory effects. Moreover, we have tested and evaluated the stability of the solutions provided against imperfections of analog components by introducing up to $10\%$ error in their corresponding coefficients, finding that the relative error reaches up to $13\%$ for relevant timescales. From these results, we have concluded the robustness and resilience of the results of the simulation provided by the methods introduced in this work. Moreover, it is noteworthy to mention that our results have been obtained with minimal architecture variations, leaving the possibility of more intrincated arrangements, whose complexity grows in complexity extremely fast, open for future studies. \\
The analog simulator can benefit from a large array of memristors, coupled in series or parallel, specially if we consider the Taylor expansion for any memductance function, which is directly related to the class of integro-differential equations that can be implemented by the circuit. The main limitation in using a large number of memristors is the error accumulation due to imperfect fabrication of the devices. In this sense, the flexibility of the analog simulator with the number of memristors increases as the device fabrication and design improves.

As a further scope, it would be interesting to consider the proposed scenarios in the context of quantum memristors~\cite{Pfeiffer2016,Shevchenko2016,Salmilehto2017,Sanz2017}. These devices are the quantized version of the memristors employed in this work, i.e. open non-Markovian controllable open quantum systems which are able to process quantum information and can be used for simulations and for implementation of quantum neural networks~\cite{Gonzalez-Raya2019_1,Gonzalez-Raya2019_2}. The intuition is that quantum entanglement could be useful in the search of quantum speed-ups for analog computers, in a similar manner as it already happens with quantum simulations.\\

\section*{Acknowledgments}
The authors would like to thank Unai Alvarez-Rodriguez and Massimiliano Di Ventra for useful discussions.
G.A.B. acknowledges support from CONICYT Doctorado Nacional 21140587 and Direccion de Postgrado USACH. J.C.R. thanks FONDECYT for support under grant No. 1140194. While M.S. and E.S. are grateful for the funding from Spanish MCIU/AEI/FEDER (PGC2018-095113-B-I00), Basque Government IT986-16, the projects QMiCS (820505) and OpenSuperQ (820363) of the EU Flagship on Quantum Technologies and the EU FET Open Grant Quromorphic (828826). This work is supported by the U.S. Department of Energy, Office of Science, Office of Advanced Scientific Computing Research (ASCR) quantum algorithm teams program, under field work proposal number ERKJ333. 

\section*{Author Contributions}
G.A.B., as the first author, has been responsible for the development of this work. M.S. suggested the seminal ideas and has contributed to the mathematical demonstrations and examples. E.S. and J.C.R. have helped to improve the ideas and results shown in the manuscript. All authors have carefully proofread the manuscript. M.S. supervised the project throughout all stages.

\section*{Additional Information}
\textbf{Competing interests}: The authors declare no financial and non-financial competing interests.
\end{document}